\documentclass[
    ,final            
  ]
  {aipproc}

\layoutstyle{6x9}

\newcommand{\bfi}[1]{\mbox{\boldmath $#1$}}
\newcommand{\Lower}[1]{\smash{\lower 1.5ex \hbox{#1}}}

\begin{document}

\title{
Study of Pentaquark and $\bfi{\Lambda}$(1405)
}


\author{
H.~Nemura
}{
  address={
Advanced Meson Science Laboratory, DRI, RIKEN,
 Wako, Saitama 351-0198, Japan 
}
}

\author{
C.~Nakamoto
}{
  address={
Suzuka National College of Technology, Suzuka, Mie 510-0294, Japan
}
}

\begin{abstract}
 We perform a five-body calculation for pentaquark $(q^4\bar{q})$ 
 state of $\Lambda(1405)$ as well as the three-body calculations for the 
 ground state baryons and the $\Lambda(1405)$,
 and two-body calculations for mesons. 
 The hamiltonian,
 which reproduces reasonably well the energies of ground state baryons 
 ($N,\Lambda,\Sigma,\Xi$ and $\Delta$) and mesons ($\pi,K,\rho,K^\ast$), 
 includes kinetic energy of semi-relativistic form,
 linear confinement potential, 
 and the simplest form of color-magnetic interaction with Gaussian 
 form factor. 
 Flavor symmetry breaking ($m_s> m_{u,d}$) is taken into account. 
 The energy calculated for $(q^4\bar{q})$ state of $\Lambda(1405)$ 
 is lower than the energy for the $(q^3)$ state. 
 The present result suggests that the $\Lambda(1405)$ is a 
 pentaquark-dominated state if the color-magnetic potential plays a
 leading role of the quark-quark and quark-antiquark interactions. 
\end{abstract}

\maketitle

\section{Introduction}

The $\Lambda(1405)$, $J^\pi={1\over 2}^-$, has the smallest mass in the 
negative parity states of the baryon spectrum. 
Several constituent quark-model studies have been mentioning that 
the contribution of $(q^4\bar{q})$ configurations could be large in the 
$\Lambda(1405)$~\cite{Brau,Helminen}. 
If this is the case, 
all of the five quarks may occupy their lowest $(0s)^5$ orbits, 
since an antiquark has the intrinsic negative parity. 
If we assume that the flavor $SU(3)$ is an exact symmetry, 
one can find that two kinds of 
the $(q^4\bar{q})$ states                      for $\Lambda(1405)$, 
 \begin{equation}
 |\Lambda(1405)\rangle=[5]_{R}[222]_{C}[222]_{F}[32]_{S},
 \end{equation}
give strongly attractive color-magnetic interactions: 
(The subscripts ($RCF$ and $S$) stand for the position coordinates, color,
flavor and spin space, respectively.)
\begin{equation}
 \left\langle \sum_{i<j} (\mathbf{\lambda}_i^C\cdot\mathbf{\lambda}_j^C)
 (\mathbf{\sigma}_i\cdot\mathbf{\sigma}_j)\right\rangle
  = \left\{
      \begin{array}{lc}
       16 & (\mbox{ for }[22]_S \mbox{ of } (q^4)), \\
       80/3 & (\mbox{ for }[31]_S \mbox{ of } (q^4)),
      \end{array}
\right.
\end{equation}
where 
$\lambda^C$ stands for the color $SU(3)$ generator, and 
$\mathbf{\sigma}$ is Pauli matrices for the spin. 
These attractive forces 
make lower the mass 
of the ($q^4\bar{q}$) state 
than that 
of the ($q^3$). 
However, 
the flavor symmetry is manifestly broken. 
A precise five-body calculation as well as the three-body calculation 
should be performed in order to clarify whether the mass of 
$(q^4\bar{q})$ state for the $\Lambda(1405)$ is still smaller than that 
of $(q^3)$ state, even if the flavor symmetry breaking is taken into 
account. 
Therefore, 
the purpose of this study is to describe a five-body calculation of 
pentaquark ($q^4\bar{q}$) state for the $\Lambda(1405)$. 

\section{Interactions and Method}

The hamiltonian is given by 
\begin{equation}
H=\sum_{i=1}^{A} \sqrt{m_i^2+\mathbf{p}_i^2}
 + \sum_{i<j} V_{ij}^{\rm (conf)} + \sum_{i<j} V_{ij}^{\rm (CM)}, 
\qquad \left(
\mbox{with}\sum_{i=1}^A \mathbf{p}_i=0
\right), 
\end{equation}
where $m_i$ and $\mathbf{p}_i$ are the mass and the momentum operator 
of the $i$-th quark (or antiquark). 
The two-body ($qq$ or $q\bar{q}$) interaction 
consists of a confinement potential and a color-magnetic potential:
\begin{equation}
V_{ij}^{\rm (conf)} = \left(-{3\over 8}\right)
\left(\lambda^C_i\cdot \lambda^C_j\right)
\left(-V_0+Cr_{ij}\right),\quad\mbox{and}
\end{equation}
\begin{equation}
V_{ij}^{\rm (CM)} = -\alpha {2\pi\over 3m_i m_j}
\left({\lambda^C_i\over 2}\cdot {\lambda^C_j\over 2}\right)
\left(\mathbf{\sigma}_i\cdot \mathbf{\sigma}_j\right)
\times {1\over \left(2\sqrt{\pi}\Lambda\right)^3}
\exp\left\{-{r_{ij}^2\over 4 \Lambda^2}\right\}, 
\end{equation}
where $r_{ij}=|\mathbf{r}_i-\mathbf{r}_j|$ is the interparticle 
coordinate. 
The delta-function in the color-magnetic potential is replaced by a
Gaussian form factor with the size parameter $\Lambda$. 
All of the parameters of the present hamiltonian are given in 
Table~\ref{parameters}. 
\begin{table}[t]
\caption{Parameters of the present model. 
}
\label{parameters}
\newcommand{\m}{\hphantom{$-$}}
\newcommand{\cc}[1]{\multicolumn{1}{c}{#1}}
\begin{tabular}{@{}cccccc}
\hline
 $m_{u,d}$ & $m_s$ & $V_0$ & $C$ & $\alpha$ & $\Lambda$ \\
\hline 
 0.34 GeV & 0.5508 GeV & 0.4534 GeV & 0.08265 (GeV)$^2$ & 1.08 & 0.204 fm \\
\hline
\end{tabular}\\[2pt]
\end{table}

The energies of various systems are calculated by the 
stochastic variational method (SVM)\cite{SVM}. %
 The trial function is given by a combination of basis functions: 
 \begin{equation}
 \Psi = \sum_{k=1}^K c_k\varphi_k,
 \qquad \mbox{with}\qquad 
 \varphi_k = {\cal A}\left\{
  G(\mathbf{x};A_k)
  \ v_k^{L_k}\ \left[Y_{L_k}(\hat{\mathbf{v}}_k)\times \chi_{S_k}
	   \right]_{JM}
  \eta_{kIM_I}\xi_{k(00)}
  \right\}.
 \end{equation}
 Here ${\cal A}$ is an antisymmetrizer acting on the identical
 particles. 
For the spin                                          $(\chi_{k})$, 
the isospin                                           $(\eta_k)$, 
and the color                                         $(\xi_k)$ 
functions, 
all possible configurations are taken into account. 
 The abbreviation $\mathbf{x}=(\mathbf{x}_1, \cdots, \mathbf{x}_{A-1})$ 
 is a set of relative coordinates. 
 For the spatial part, the basis function is constructed by the 
 correlated Gaussian (CG), $G(\mathbf{x};A_k)$,
 multiplied by the orbital angular momentum part. 
 The CG is given by 
 \begin{equation}
 G(\mathbf{x};A_k)
 =\exp\left\{-{1\over 2}\sum_{i<j}^A\alpha_{kij}
 (\mathbf{r}_i-\mathbf{r}_j)^2
 \right\}
 =\exp\left\{-{1\over 2}\sum_{i,j=1}^{A-1}A_{kij}
 \mathbf{x}_i\cdot\mathbf{x}_j
 \right\}.
 \end{equation}
The orbital angular momentum part, which is needed to the $(q^3)$ model 
of $\Lambda(1405)$, is expressed by the global vector representation. 
The global vector, $\mathbf{v}_k$, is given by a linear combination of 
the relative coordinates: 
\begin{equation}
 \mathbf{v}_k = \sum_{i=1}^{A-1}(u_k)_i \mathbf{x}_i.
\end{equation}
The $A_k$ and $u_k$ are sets of nonlinear parameters which characterize 
the spatial part of the basis function. 
The variational parameters are optimized by a stochastic procedure. 
The SVM                                     with the above CG basis 
produces accurate solutions. 
The reader is referred to Refs.\cite{abinitio,fullycoupl} %
for details and recent applications.

\section{Results and Discussion}

\begin{table}[t]
\caption{
 Energies of baryons and mesons, given in units of MeV. 
 The energies of the three-body model and the five-body model for 
 $\Lambda(1405)$ are also given. 
}
\label{results}
\newcommand{\m}{\hphantom{$-$}}
\newcommand{\cc}[1]{\multicolumn{1}{c}{#1}}
\begin{tabular}{@{}lcccccccccccccc}
\hline
 & \multicolumn{5}{c}{Baryons} & & \multicolumn{4}{c}{Mesons} & & 
 \multicolumn{2}{c}{$\Lambda(1405)$}\\
 \cline{2-6} \cline{8-11} \cline{13-14}
 & $N$ & $\Delta$ & $\Lambda$ & $\Sigma$ & $\Xi$ & & 
 $\pi$ & $K$ & $\rho$ & $K^\ast$ & & $(q^3)$ & $(q^4\bar{q})$ \\
\hline
 Calc. & 949 & 1266 & 1116 & 1208 & 1336 & & 141 & 543 & 771 & 907 & & 1405 & 1292 \\
 Expt. & 939 & 1232 & 1116 & 1193 & 1318 & & 137 & 496 & 776 & 892 & & \multicolumn{2}{c}{$1406\pm 4$} \\
\hline
\end{tabular}\\[2pt]
\end{table}
Table~\ref{results} lists the energies of three-body calculations 
(ground state baryons) and of two-body calculations (mesons). 
All of the calculated energies reasonably well reproduce the
experimental values. 
The table also shows the energies of the three-body model and 
of the five-body model for the $\Lambda(1405)$. 
The present ($q^3$) model for the $\Lambda(1405)$ happens to 
reproduce the correct mass. 
However, this is not the point in the present study. 

The remarkable result is seen in the five-body model of the 
$\Lambda(1405)$. 
The energy calculated for the $(q^4\bar{q})$ state is lower than that 
for the $(q^3)$ state. 
Therefore, the present result suggests that the $\Lambda(1405)$ is 
a pentaquark-dominated state if the color-magnetic potential 
plays a leading role of the $q-q$ and $q-\bar{q}$ interactions. 
In the present model, 
the $(q^4\bar{q})$ state is a bound state 
since 
the energy 
obtained for the $(q^4\bar{q})$ state 
is lower than both the $\pi+\Sigma$ and the $\bar{K}+N$ thresholds, 
which are calculated to be 1348 MeV and 1492 MeV, respectively. 
More realistic model, e.g., taking account of 
effective meson-exchange force 
or 
coupling potential between $(q^3)$ and $(q^4\bar{q})$, 
will be described in future publication. 
An attempt along this line has been made 
in Ref.~\cite{Nakamoto}.

\begin{theacknowledgments}
H.~N. is supported by the Special Postdoctoral Researchers Program 
at RIKEN. 
This study was supported by Grants-in-Aid for Young Scientists (B) 
(No.~17740174 and No.~15740161) from the Japan Society for the 
Promotion of Science (JSPS). 
\end{theacknowledgments}

\bibliographystyle{aipproc}   

\bibliography{sample}

\begin{thebibliography}{9}


\bibitem{Brau}
F.~Brau, C.~Semay, and B.~Silvestre-Brac,
\emph{Phys. Rev. C} \textbf{66}, 055202 (2002). 

\bibitem{Helminen}
C.~Helminen, and D.~O.~Riska, \emph{Nucl. Phys. A} \textbf{699}, 624- (2002). 


\bibitem{SVM} Y.~Suzuki, and K.~Varga, 
\emph{Stochastic Variational Approach to 
Quantum-Mechanical Few-Body Problems}, Lecture Notes in 
Physics, Vol. m54 (Springer-Verlag, Berlin Heidelberg, 1998). 


\bibitem{abinitio}
H.~Nemura, Y.~Akaishi, and Y.~Suzuki, 
\emph{Phys. Rev. Lett.} \textbf{89}, 142504 (2002). 

\bibitem{fullycoupl}
H.~Nemura, S.~Shinmura, Y.~Akaishi and Khin Swe Myint, %
\emph{Phys. Rev. Lett.} \textbf{94}, 202502 (2005). 

\bibitem{Nakamoto} C.~Nakamoto, and H.~Nemura,
	in these proceedings. 


\end{thebibliography}

\IfFileExists{\jobname.bbl}{}
 {\typeout{}
  \typeout{******************************************}
  \typeout{** Please run "bibtex \jobname" to optain}
  \typeout{** the bibliography and then re-run LaTeX}
  \typeout{** twice to fix the references!}
  \typeout{******************************************}
  \typeout{}
 }

\end{document}